\documentclass[a4paper,reqno,superscriptaddress,nofootinbib, pra, aps,11pt]{revtex4-2}
\usepackage[centertags]{amsmath}
\usepackage{amsfonts}
\usepackage{amssymb}
\usepackage{amsthm}
\usepackage{newlfont}
\usepackage{stmaryrd}
\usepackage{mathrsfs}
\usepackage{mathtools}
\usepackage{euscript}
\usepackage{graphicx}
\usepackage{enumerate}
\usepackage{todonotes}

\usepackage{orcidlink}

\usepackage{amsmath}

\usepackage{tikz}
\usepackage{pgf}
\usetikzlibrary{positioning,fit,calc}
\usetikzlibrary{arrows,automata}
\usepackage{wrapfig}
\usepackage{amscd}
\usepackage{cancel}
\usepackage{hhline}
\usepackage{import}

\usepackage{changes}

\theoremstyle{plain}
\newtheorem{theorem}{Theorem}[section]

\theoremstyle{definition}

\theoremstyle{remark}

\let\ve=\varepsilon

\newcommand{\be}{\begin{equation}}
\newcommand{\en}{\end{equation}}

\newcommand{\opunit}{\text{1}\kern-0.22em\text{l}}

\newcommand{\id}{\textrm{d}}

\SetSymbolFont{stmry}{bold}{U}{stmry}{m}{n}

\DeclareMathAlphabet{\mathpzc}{OT1}{pzc}{m}{it}

\let\oldsqrt\sqrt
\def\sqrt{\mathpalette\DHLhksqrt}
\def\DHLhksqrt#1#2{%
	\setbox0=\hbox{$#1\oldsqrt{#2\,}$}\dimen0=\ht0
	\advance\dimen0-0.2\ht0
	\setbox2=\hbox{\vrule height\ht0 depth -\dimen0}%
	{\box0\lower0.4pt\box2}}

\let\ve=\varepsilon

\let\be=\beta

\DeclareMathAlphabet{\mathpzc}{OT1}{pzc}{m}{it}

\def\bea{\begin{eqnarray}}
\def\eea{\end{eqnarray}}
\def\ba{\begin{array}}
	\def\ea{\end{array}}

\usepackage{url}
\usepackage{hyperref}

\usepackage{titlesec}

\hypersetup{colorlinks=true}
\hypersetup{linkcolor=black}

\usepackage{float}

\usepackage{silence}
\usepackage{mdframed}

\definecolor{ring color}{rgb}{0.31,0.31,0.34}
\definecolor{border color}{rgb}{0.85,0.8627,0.839}

\newcommand{\married}{%
\begin{tikzpicture}[x=1ex, y=1ex, scale=0.5, baseline=-.6ex]
\begin{scope}
\clip (-2,-.2) rectangle ++(5.5,2.2);
\path[draw=border color,line width=.08ex,
     fill=ring color,even odd rule]
     (0, 0) circle (2ex)
     (0, 0) ellipse (1.5ex and 1.3ex);
\path[draw=border color,line width=.08ex,
     fill=ring color,even odd rule]
     (1.5, 0) circle (2ex)
     (1.5, 0) ellipse (1.5ex and 1.3ex);
\end{scope}
\begin{scope}
\clip (-2,-2) rectangle ++(5.5,2.2);
\path[draw=border color,line width=.08ex,
     fill=ring color,even odd rule]
     (1.5, 0) circle (2ex)
     (1.5, 0) ellipse (1.5ex and 1.3ex);
\path[draw=border color,line width=.08ex,
     fill=ring color,even odd rule]
     (0, 0) circle (2ex)
     (0, 0) ellipse (1.5ex and 1.3ex);
\end{scope}
\end{tikzpicture}%
}

\begin{document}
\title{First part of Clausius heat theorem in terms of Noether's theorem}
\author{Aaron Beyen \orcidlink{0000-0002-4341-7661}}
\author{Christian Maes \orcidlink{0000-0002-0188-697X}}

\affiliation{Department of Physics and Astronomy, KU Leuven, 3001 Leuven, Belgium \\ 
{\normalfont\texttt{aaron.beyen, christian.maes@kuleuven.be}}
}

\begin{abstract}
After Helmholtz, the mechanical foundation of thermodynamics included the First Law $\id E = \delta Q + \delta W$, and the first part of the Clausius heat theorem $\delta Q^\text{rev}/T = \id S$. The resulting invariance of the entropy $S$ for quasistatic changes in thermally isolated systems invites a connection with Noether's theorem (only established later). In this quest, we continue an idea, first brought up by Wald in black hole thermodynamics and by Sasa {\it et al.} in various contexts.  We follow both Lagrangian and Hamiltonian frameworks, and emphasize the role of Killing equations for deriving a First Law for thermodynamically consistent trajectories, to end up with an expression of ``heat over temperature'' as an exact differential of a Noether charge.
\end{abstract}
\maketitle

entropy, Killing equations.

\section{Introduction}
The operational definition of entropy\footnote{attributed to Rudolf Clausius, \cite{clausius1865}} for macroscopic equilibrium systems  considers a reversible process of heating a substance in many small quasistatic steps for which
\begin{equation}\label{ds}
\id S = \frac{\delta Q^\text{rev}}{T}
\end{equation}
connecting the entropy increase $\id S$ during a step to the absorbed heat $\delta Q^\text{rev}$ from a bath at instantaneous temperature $T$. That the right-hand side of \eqref{ds} is indeed an exact differential is a result known as the (first part of the) Clausius heat theorem \cite{clausius1865}.
For the \textit{microscopic} derivation of the \textit{macroscopic} property \eqref{ds}, controversies remain  about   the assumptions and the required level of description, and  even  about  the very meaning of heat and temperature \cite{Loschmidt, noneqtemperature, microclausius}. At any rate,  Ludwig Boltzmann and Rudolf Clausius still had a more mechanical picture in mind, \cite{Boltzmann_1866, clausius1871, clausius21871}, and started their search toward entropy, triggered by the bicentennial work of Sadi Carnot \cite{carnot1824reflexions}, by analyzing one-dimensional periodic dynamical systems. Their ``ensembles'' (to borrow a word attributed to Josiah Gibbs, much later) were still based on time--averages where \textit{e.g.} the temperature corresponds to twice the average kinetic energy. Their result is therefore related to a theorem by Hermann von Helmholtz, \cite{Helmholtza, Helmholtzb}, and implies that the so-called Helmholtz entropy is an adiabatic invariant (in the sense of classical mechanics\footnote{We recall that the term `adiabatic’ in mechanics differs from the thermodynamic use where it stands for `thermally isolated’. The mechanical definition corresponds to a very slow transformation during which the adiabatic invariant remains constant \cite{landau1976mechanics}. For a conservative mechanical system at energy $E$, we call a function $I(E,\lambda)$ an ``adiabatic invariant'' if in the quasistatic limit where  $\lambda=\lambda(\ve t)$ is time-dependent with rate  $\ve \to  0$,   $I(E(t),\lambda(t))\rightarrow $ constant. }), \cite{Campisi_2004, Campisi_2010}. Only later, and mostly emphasized by Boltzmann, the time-averages were replaced by Gibbs statistical averages, as in the microcanonical ensemble, and the ``heat over temperature'' was related with the (Boltzmann) volume entropy ($\log$ of the phase-space volume enclosed by the energy surface).\\

Recent years have seen a revival of a more (purely) mechanical foundation of Clausius' Heat Theorem.  We do not have  in mind so much  the emphasis on smooth dynamical systems or the chaotic hypothesis, \cite{microclausius,ruelle,Gal2,thermo_kepler,kurc}. Rather, we mean developments through the work of Wald \textit{et al}. \cite{wald_original, wald2}, Sasa \textit{et al} \cite{Sasa_original, Sasa_quantum, langevin_noether}  and Bravetti \textit{et al} \cite{contactgeometry}. They show how entropy can sometimes be realized as a Noether charge for classical, quantum, Langevin, black hole systems  and from a contact geometry viewpoint, thereby connecting the Clausius to the Noether theorem, \cite{Noether1918}.  
However, the relation between these approaches and results has largely remained unclear and underexplored. Our approach starts in the style of Wald \cite{wald_original}, by deriving the First Law from the quasisymmetry of the action under some types of transformations; from there we understand how to identify ``heat over temperature'' which is then identified with a Noether charge, to reach the results by Sasa {\it et al} in \cite{Sasa_original}.  We add therefore in the present paper a connection via the Killing equations, natural indeed for the context of black hole geometry and Noether's invariance in classical mechanical systems.    New is also the analysis in both the Lagrangian and Hamiltonian formalism and a different, equivalent symmetry transformation allowing connections with \cite{langevin_noether, Beyengradientflow, beyengeneric}. \\
 Furthermore, introducing Noether's theorem into statistical physics has been addressed in recent years, see {\it e.g.}, \cite{Hermann_2021, Hermann_2022, 2Hermann2022, 3Hermann_2022, Herman_2023, Hermann_2024, Robitschko2024, 2Herman_2024}. \\ 

Connecting the invariance of the entropy under quasistatic adiabatic transformations with Noether's theorem may be surprising at first since the latter involves a {\it continuous} symmetry (transformation), obtained \textit{e.g.} by shifting time or space, while entropy is deeply and uniquely associated with time-reversal \cite{maes2002timereversalentropy, Maes_2012}, a \textit{discrete} transformation. 
Our intuition is that the discrete symmetry may be derived from a continuous one in a higher-dimensional space, similar to how a (discrete) reflection can be viewed as a (continuous) rotation in higher dimensions. There is not only the coordinate space or even the symplectic space of the Hamiltonian; there is also the space of forces  that act on the system and can drive the system  one way or the other.  For an inspiring analogy, we mention the conservation of optical helicity \cite{afanasiev1996helicity} and the zilch \cite{lipkin} in (source-free) electromagnetism, related to the discrete electromagnetic duality $(\vec{E}, \vec{B}) \to (\vec{B}, -\vec{E})$. In \cite{zilch2, ELBISTAN20181897, Bliokh_2013}, one adds a vector potential $C^\mu$ on equal footing with $A^\mu$ to the free Maxwell Lagrangian density $\cal L$, and shows how the conservation laws follow from Noether's theorem under continuous `rotations' between $A^\mu, C^\mu$. As a matter of fact, as we stipulate around equation \eqref{conserved quantity hamiltonian}, whenever there is a conserved quantity which is sufficiently smooth as a function of the dynamical variables, there exists a continuous symmetry and corresponding Noether charge which equals that quantity.  It even allows to conceive related but different continuous symmetries, which removes some of the difficulty in interpreting such things as ``inhomogeneous time-translation''. For instance, the intuition of a `rotation' between $q,p$ (akin to $A^\mu, C^\mu$) and its projection in the Lagrangian formalism will appear.\\

\underline{Plan of the paper:} 
We start in Section \ref{section deterministic systems} by introducing the mechanical setup and recalling the essential steps of the Noether theorem for a general symmetry transformation.\\
The additional ingredient in Section \ref{sec3} is to apply a quasistatic protocol in the external parameters and an averaging procedure. The central result is exposed in Theorem \ref{cenr}. Following the ideas of \cite{Sasa_original}, it is shown
how a quasisymmetry of the action leads to an exact differential for the Noether charge $\mathcal{N}$ which can be identified as ``heat over temperature,'' depending on a thermodynamic averaging procedure. 
We proceed via the First Law to identify that heat.\\
Section \ref{rem} and the proofs in Section \ref{section interpretation} are essential for interpreting other novelties of our approach.  We elaborate on the connection with the work in \cite{wald_original}, and we present various other but connected symmetry transformations for which the result of the Theorem \ref{cenr} is verified.\\
We end in Section \ref{section examples} with well-known examples of (one-dimensional) periodic systems, which are worked out in detail to explicitly clarify the relation between (Helmholtz and volume) entropy and Noether charge.

\section{Mechanical heat theorem}\label{section deterministic systems}
The first part of the Clausius heat theorem \eqref{ds} is an important element in the (early) mechanical foundations of thermodynamics.  We refer to the expositions \cite{Campisi_2010,Campisi_2004, thermo_kepler,Gal2} where not only much of the original work is discussed but also connected with future developments. The present paper, as previous ones like \cite{Sasa_original}, results from the wish to connect the principle of adiabatic invariance of the entropy, which was made central by Paul Hertz\footnote{Hertz showed that the volume entropy (phase-space volume enclosed by the energy surface) is an adiabatic invariant.}, \cite{hertz1910}, with the Noether theorem. However, the perspective is somewhat different from \cite{Sasa_original} as we prefer to proceed via the First Law.  In other words, we first derive a First Law to characterize the heat and then proceed to show that this ``heat over temperature'' is an exact differential. 

\subsection{Mechanical First Law}
We consider a mechanical system with (generalized) coordinates $q$.
For a given Lagrangian $\cal L=\cal L(q,\dot q; \lambda)$, we consider the action 
\begin{equation}\label{action in summary}
    \mathcal{A} = \int_{t_1}^{t_2} \id t \ \cal L(q(t),\dot{q}(t); \lambda(t))
\end{equation}
where $ \left( q(t), t\in [t_1,t_2] \right)$ is a (possible mechanical) trajectory, and
where $\lambda(t)$ is an external control (protocol)\footnote{Here it just means that the Lagrangian (and Hamiltonian) is time-dependent, but we prefer to suggest a more physical interpretation that becomes convenient for the quasistatic limit later.}  for the variation of parameters $\lambda$.  We use mostly one-dimensional notation for simpllicity.   The same steps can be performed for $N$ particle systems (as was one of the examples in \cite{Sasa_original}).\\

Next,  we introduce the energy function 
\begin{equation}\label{energy function summary}
E =    E(q, \dot{q}; \lambda) = \frac{\partial \cal L}{\partial \dot{q}} \ \dot{q} - \cal L
\end{equation}
 which is assumed to be constant when $\lambda$ does not change in time, {\it i.e.}, the system is conservative. When a time-dependence $\lambda = \lambda(t)$ is introduced, the energy changes as    
\begin{align}\label{firsteuler}
    \frac{\id E}{\id t} - \frac{\partial E}{\partial \lambda} \frac{\id \lambda}{\id t} &= \frac{\id }{\id t} \left(\frac{\partial \cal L}{\partial \dot{q}} \dot{q} - \cal L\right) - \frac{\partial}{\partial \lambda} \left(\frac{\partial \cal L}{\partial \dot{q}} \dot{q} - \cal L \right) \frac{\id \lambda}{\id t} \nonumber \\
    & = \frac{\id}{\id t} \left(\frac{\partial \cal L}{\partial \dot{q}} \right) \dot{q} + \frac{\partial \cal L}{\partial \dot{q}} \ddot{q}- \frac{\partial \cal L}{\partial q} \dot{q} - \frac{\partial \cal L}{\partial \dot{q}} \ddot{q} - \frac{\partial \cal L}{\partial \lambda} \frac{\id \lambda}{\id t}  - \frac{\partial}{\partial \lambda} \left(\frac{\partial \cal L}{\partial \dot{q}} \dot{q} - \cal L \right) \frac{\id \lambda}{\id t} \nonumber \\
    &= \left(\frac{\id}{\id t} \left(\frac{\partial \cal L}{\partial \dot{q}} \right) - \frac{\partial \cal L}{\partial q} \right) \dot{q} - \frac{\partial}{\partial \lambda} \left( \frac{\partial \cal L}{\partial \dot{q}} \dot{q} \right) \frac{\id \lambda}{\id t} \nonumber \\
    & = -\mathscr{E} \dot{q} - \frac{\partial}{\partial \lambda} \left( \frac{\partial \cal L}{\partial \dot{q}} \dot{q} \right) \frac{\id \lambda}{\id t} 
\end{align}
where  we pin the Euler-Lagrange functional as
\begin{equation}\label{EL functional}
    \mathscr{E} = \frac{\partial \cal L}{\partial q} - \frac{\id}{\id t} \Big( \frac{\partial \cal L}{\partial \dot{q}} \Big)
\end{equation}
The identity \eqref{firsteuler} can be written in a form resembling the First Law: combining \eqref{energy function summary} with \eqref{firsteuler}, we obtain
\begin{eqnarray}\label{mechanical first law lagrangian}
    \id E &=& \delta Q + \delta W, \qquad\text{ for }\\
    \delta W &=& -\frac{\partial}{\partial \lambda} \left( \frac{\partial \cal L}{\partial \dot{q}} \dot{q} - E \right) \ \id \lambda = -\frac{\partial \cal L}{\partial \lambda} \id \lambda, \qquad \delta Q = - \mathscr{E} \ \id q\nonumber
\end{eqnarray}
with conjugate force $F = \partial_\lambda \cal L$ in the work $\delta W$. Note in particular that the heat $\delta Q$ vanishes \textit{on shell}\footnote{Here, and in the rest of the paper, working ``on shell'' refers to assuming the equations of motion (Euler-Lagrange or Hamilton's equations) are satisfied. For example, $\mathscr{E} = 0$ and $\delta Q=0$ on shell. Whenever there is a nonconservative force ({\it i.e.} friction forces) in the Euler-Lagrange equations \eqref{EL functional}
\begin{equation*}
    \mathscr{E} = \frac{\partial \cal L}{\partial q} - \frac{\id}{\id t} \Big( \frac{\partial \cal L}{\partial \dot{q}} \Big) = - F
\end{equation*}
then $\delta Q = F \ \id q$ on shell.}. The relation \eqref{mechanical first law lagrangian} is a mechanical First Law in Lagrangian form where the differential $\id E = \frac{\id E}{\id t} \id t $ refers to the time-derivative for the energy function \eqref{energy function summary} depending on the coordinates and the protocol but not using the equations of motion.\\

Under the usual assumptions, the Hamiltonian is related to the Lagrangian through a Legendre transformation 
\begin{equation*}
    \cal H(q, p, \lambda) = \sup_{v} \left(p v - \cal L(q,v, \lambda) \right), \qquad \cal L(q, \dot{q}, \lambda) = \sup_{p} \left(p \dot{q} - \cal H(q,p, \lambda) \right)
\end{equation*}
with $p$ the momentum, leading to the action
\begin{equation}\label{hamiltonian action classical}
    \cal A = \int_{t_1}^{t_2} \id t \ \left[p(t) \ \dot{q}(t) - \cal H(q(t),p(t), \lambda(t)) \right]
\end{equation}
where a trajectory is now a path in phase space specified by the couple $(q(t), p(t)), t \in [t_1,t_2]$. The identity \eqref{firsteuler} gets replaced by
\begin{equation}\label{dhdt}
    \frac{\id \cal H}{\id t} - \frac{\partial \cal H}{\partial \lambda} \frac{\id \lambda}{\id t} = \frac{\partial \cal H}{\partial q} \dot{q} + \frac{\partial \cal H}{\partial p} \dot{p} =  \frac{\partial \cal H}{\partial q} \mathscr{E}_q + \frac{\partial \cal H}{\partial p} \mathscr{E}_p
\end{equation}
with
\begin{equation*}
    \mathscr{E}_q = \dot{q} - \frac{\partial \cal H}{\partial p}, \qquad \mathscr{E}_p = \dot{p} + \frac{\partial \cal H}{\partial q} 
\end{equation*}
Note that we do not get a derivative $\partial_\lambda p$ as in \eqref{firsteuler} since the momentum is an independent parameter now. The mechanical First Law \eqref{mechanical first law lagrangian} becomes
\begin{eqnarray}
    \id \cal H &=& \delta Q + \delta W, \qquad  \text{ with } \\
    \delta W &=& \frac{\partial \cal H}{\partial \lambda} \ \id \lambda, \qquad \delta Q = \frac{\partial \cal H}{\partial q} \id q + \frac{\partial \cal H}{\partial p} \id p\nonumber
\end{eqnarray}
with generalized force $F = - \partial_\lambda \cal H$, for which
\begin{equation}\label{partial derivative H,L}
    \left( \frac{\partial \cal H}{\partial \lambda} \right)_{q,p} = -  \left(\frac{\partial \cal L}{\partial \lambda} \right)_{q, \dot{q}}
\end{equation}

\subsection{Noether's  theorem}\label{appendix noether theorem}
For self-consistency and choice of notation, we recall the formal setup of Noether's theorem.\\
Given a mechanical trajectory, we consider the symmetry transformations
\begin{align}
    \text{(Lagrangian)} :  t \to t' = t + \eta \,\xi(q(t), \dot{q}(t), t) \qquad &q \to q'(t') = q(t) +\eta \ \delta q(q(t), \dot{q}(t), t) \label{general lagrangian transformation}
    \\
\text{(Hamiltonian)} :   t \to t' = t + \eta \ \xi(q(t), p(t), t), \qquad &q \to q'(t') = q(t) +\eta \ \delta q(q(t), p(t), t) \label{general hamiltonian transformation} \\ 
&p \to p'(t') = p(t) +\eta \ \delta p(q(t), p(t), t) \nonumber
\end{align}
with  ``infinitesimal'' parameter $\eta$.
Following \textit{e.g.} \cite{laghamconnection}, a variational calculation gives the change in action $\cal A'-\cal A =\delta \cal A$ which in the Lagrangian, respectively, in the Hamiltonian framework reads as
\begin{align}
   &\delta \mathcal{A} = \eta \int_{t_1}^{t_2} \id t \ \left[ (\delta q - \xi \dot{q} ) \left(\frac{\partial \cal L}{\partial q} - \frac{\id}{\id t} \left( \frac{\partial \cal L}{\partial \dot{q}} \right) \right) \, - \frac{\id}{\id t} \left( \xi\left[\frac{\partial \cal L}{\partial \dot{q}} \,\dot{q} - \cal L\right] - \frac{\partial \cal L}{\partial \dot{q}} \delta q  \right) \right] \label{change action lagrangian general} \\
    &\delta \mathcal{A} = \eta \int_{t_1}^{t_2} \id t \ \Bigg[(\delta p - \xi \dot{p} ) \left(\dot{q} - \frac{\partial \cal H}{\partial p}  \right) - (\delta q-\xi \dot{q}) \, \left(\dot{p} + \frac{\partial \cal H}{\partial q} \right) - \frac{\id}{\id t} \Big( \xi \cal H - p \delta q \Big) \Bigg]  \label{change action hamiltonian general}
\end{align}
In the above, we have assumed a fixed protocol since it is not a degree of freedom, {\it i.e.},  $\lambda'(t) = \lambda(t)$, as in \cite{Sasa_original}. That is mainly a matter of notation, to say that the Lagrangian should be seen as a function of the coordinates $q(t), \dot{q}(t), t$. It does not mean that the protocol itself is invariant under spacetime transformations.\\
Suppose now that \eqref{general lagrangian transformation}--\eqref{general hamiltonian transformation}
leave the action invariant in the sense of a quasisymmetry, \cite{brown2021symmetries, Trautman, Bessel1921, sarlet}, {\it i.e.}, there exists a function $\Psi =  \Psi(q(t), \dot{q}(t),t)$  such that\footnote{Noether's theorem allows for $\Psi \neq 0$.  For instance, boundary terms like $\int_{t_{1}}^{t_2} \id t \ \frac{\id \Psi}{\id t}$ appear in Noether's theorem  when studying the Galilean invariance of classical mechanics or the conservation of the Laplace-Runge-Lenz vector \cite{laghamconnection, Marinho_2009}.} 
\begin{equation}\label{psi hamiltonian}
    \delta \mathcal{A} = \eta \int_{t_{1}}^{t_2} \id t \ \frac{\id \Psi}{\id t}
\end{equation}
Then, equating \eqref{change action lagrangian general}--\eqref{change action hamiltonian general} to \eqref{psi hamiltonian} for all $t_1,t_2$, leads to the identities
\begin{align}
     &\text{(Lagrangian)}:  \frac{\id \cal N}{\id t} = (\delta q - \xi \dot{q} ) \left(\frac{\partial \cal L}{\partial q} - \frac{\id}{\id t} \left( \frac{\partial \cal L}{\partial \dot{q}} \right) \right) \label{off shell lagrangian} \\
     & \text{(Hamiltonian)} : \frac{\id \cal N}{\id t} =  (\delta p - \xi \dot{p} ) \left(\dot{q} - \frac{\partial \cal H}{\partial p}  \right) - ( \delta q -\xi \dot{q}) \, \left(\dot{p} + \frac{\partial \cal H}{\partial q} \right) \label{off shell hamiltonian}
\end{align}
where the Noether charge $\cal N$ equals
\begin{align}
    \text{(Lagrangian)} &:  \cal N = \Psi + \left( \frac{\partial \cal L}{\partial \dot{q}} \dot{q} - \cal L \right) \xi - \frac{\partial \cal L}{\partial \dot{q}} \delta q = \Psi + E(q, \dot{q},t) \xi - p(q, \dot{q},t) \delta q
    \label{general noether charge lagrangian} \\
    \text{(Hamiltonian)} &:  \cal N = \Psi +  \cal H(p,q,t) \xi - p \delta q \label{general noether charge hamiltonian}
\end{align}
On shell, the right-hand sides in \eqref{off shell lagrangian}--\eqref{off shell hamiltonian} vanish, making the Noether charge a constant of motion, {\it i.e.}, $\id \cal N/ \id t = 0$. For the standard examples: the conservation of momentum (when the  Lagrangian/Hamiltonian is translational invariant $\frac{\partial \cal L}{\partial q} = \frac{\partial \cal H}{\partial q} = 0$) is recovered for $\Psi = 0, \xi = 0, \delta q = $ constant, and conservation of energy (when the  Lagrangian/Hamiltonian is time-independent $\frac{\partial \cal L}{\partial t} = \frac{\partial \cal H}{\partial t} = 0$, $\lambda = $ constant in our mechanical setup) for $\Psi = 0, \delta q = 0, \xi = $ constant. \\
 Coming back to the quasisymmetry \eqref{psi hamiltonian} of the action, we note that the examples in Section \ref{section examples} show that $\Psi = \cal N - \xi E$ does indeed not need to vanish; one requires a quasisymmetry to obtain entropy as a Noether charge. 

\subsection{Remarks I}\label{rem1}
\begin{enumerate}
    \item
The assumption \eqref{psi hamiltonian} is obviously highly nontrivial.  A symmetry transformation \eqref{general lagrangian transformation} of a given action/Lagrangian only leads to a Noether charge as in \eqref{general noether charge lagrangian}--\eqref{general noether charge hamiltonian} when $\xi, \delta q$ satisfy a set of equations which can be recast in the form of (generalized) Killing equations\footnote{
Our central equation (31) is a rewriting of them in terms of the energy function $E$. }, \cite{killingequationclassical, orbits, sarlet}:\\ Given an infinitesimal transformation \eqref{general lagrangian transformation}, the change in the action can  be written as
\begin{equation*}
    \delta \mathcal{A} = \eta \int_{t_1}^{t_2} \id t \Bigg[  \frac{\partial \mathcal{L}}{\partial t} \xi + \frac{\partial \mathcal{L}}{\partial q} \delta q + \frac{\partial \mathcal{L}}{\partial \dot{q}} \left( \frac{\id (\delta q)}{\id t} - \dot{q} \frac{\id \xi}{\id t} \right) + \mathcal{L} \frac{\id \xi}{\id t} \Bigg]
\end{equation*}
and leads to a quasisymmetry \eqref{psi hamiltonian} only when
\begin{equation}\label{noether-bessel-hagen}
    \frac{\partial \mathcal{L}}{\partial t} \xi + \frac{\partial \mathcal{L}}{\partial q} \delta q + \frac{\partial \mathcal{L}}{\partial \dot{q}} \left( \frac{\id (\delta q)}{\id t} - \dot{q} \frac{\id \xi}{\id t} \right) + \mathcal{L} \frac{\id \xi}{\id t}=\frac{\id \Psi}{\id t}
\end{equation}
which is the Noether--Bessel-Hagen equation, \cite{Bessel1921, baker2021, noetherphilosophy}. For $\delta q = 0$ and $\cal L = \cal L(q(t), \dot{q}(t), \lambda(t))$, it can be rewritten in the form 
\begin{equation}\label{Bessel-Hagen with energy}
 \xi \frac{\id E}{\id t} +  \xi \frac{\partial \cal L}{\partial \lambda} \frac{\id \lambda}{\id t}  = \frac{\id \Psi}{\id t} + \frac{\id \xi}{\id t} \left( \frac{\partial \cal L}{\partial \dot{q}} \dot{q} - \cal L \right) + \xi \frac{\id E}{\id t} = \frac{\id}{\id t} \left(\Psi + \xi E \right) = \frac{\id \cal N}{\id t}
\end{equation}
where we added the term $\xi \ \id E/\id t$ at both sides and reintroduced the Noether charge $\cal N$, \eqref{general noether charge lagrangian}. This equation also follows from combining \eqref{firsteuler}--\eqref{off shell lagrangian}. It is essential for what follows, as it has the form of the (mechanical) First Law. \\
We can also make a connection with the Killing equation from General Relativity. 
Writing out the time-derivatives in \eqref{noether-bessel-hagen}, we obtain 
\begin{align*}
   & \frac{\partial \mathcal{L}}{\partial t} \xi + \frac{\partial \mathcal{L}}{\partial q} \delta q + \frac{\partial \mathcal{L}}{\partial \dot{q}} \left[\frac{\partial (\delta q)}{\partial t} + \frac{\partial (\delta q)}{\partial q} \dot{q} + \frac{\partial (\delta q)}{\partial \dot{q}} \ddot{q} - \dot{q}\left( \frac{\partial \xi}{\partial t} + \frac{\partial \xi}{\partial q} \dot{q} + \frac{\partial \xi}{\partial \dot{q}} \ddot{q}\right)  \right] \\
    & + \mathcal{L} \left(\frac{\partial \xi}{\partial t} + \frac{\partial \xi}{\partial q} \dot{q} + \frac{\partial \xi}{\partial \dot{q}} \ddot{q} \right) = \frac{\partial \Psi}{\partial t} + \frac{\partial \Psi}{\partial q} \dot{q} + \frac{\partial \Psi}{\partial \dot{q}} \ddot{q} \nonumber
\end{align*}
Since $\xi, \delta q$ only depend on $(t, q, \dot{q})$ (and thus not on $\ddot{q}$), the coefficient of $\ddot{q}$ has to vanish separately, leading to two independent partial differential equations for the unknown $\xi, \delta q$:
\begin{align}
    & \mathcal{L} \frac{\partial \xi}{\partial \dot{q}} + \frac{\partial \mathcal{L}}{\partial \dot{q}} \left(\frac{\partial(\delta q)}{\partial \dot{q}} - \frac{\partial \xi}{\partial \dot{q}} \dot{q} \right) = \frac{\partial \Psi}{\partial \dot{q}} \label{generalised killing eq most general 1} \\
    &\frac{\partial \mathcal{L}}{\partial t} \xi + \frac{\partial \mathcal{L}}{\partial q} \delta q + \frac{\partial \mathcal{L}}{\partial \dot{q}} \left[ \frac{\partial(\delta q)}{\partial t} + \frac{\partial(\delta q)}{\partial q} \dot{q} - \dot{q} \left(\frac{\partial \xi}{\partial t} + \frac{\partial \xi}{\partial q} \dot{q} \right) \right] = \frac{\partial \Psi}{\partial t} + \frac{\partial \Psi}{\partial q} \dot{q} \label{generalised killing eq most general 2}
\end{align}
These equations represent the necessary and sufficient conditions for the action integral to be a quasisymmetry under the infinitesimal transformation with generators $\xi, \delta q$.  Consider now a system of free or holonomic constrained mass points with Lagrangian
\begin{equation*}
    \mathcal{L}(q, \dot{q}) = \frac{1}{2} g_{kl}(q) \dot{q}^k \dot{q}^l
\end{equation*}
where $g_{kl} = g_{lk}$ is the fundamental metric tensor, under the transformation
\begin{equation*}
   t \to t' = t \qquad q'^k(t) \to q'^k(t) = q^k(t) + \eta \delta q^k(q(t))
\end{equation*}
 Then, \eqref{generalised killing eq most general 1} is trivially satisfied for $\Psi = 0$ (exact invariance), while \eqref{generalised killing eq most general 2} becomes
\begin{equation*}
   \left( \frac{\partial g_{il}}{\partial q^k} \delta q^k + g_{ik} \frac{\partial(\delta q^k)}{\partial q^l} + g_{kl} \frac{\partial(\delta q^k)}{\partial q^i} \right)\dot{q}^i \dot{q}^l = 0
\end{equation*}
Since $\dot{q}^j \dot{q}^k $ is arbitrary, its coefficient must vanish
\begin{equation*}
    0 =  \frac{\partial g_{il}}{\partial q^k} \delta q^k + g_{ik} \frac{\partial(\delta q^k)}{\partial q^l} + g_{kl} \frac{\partial(\delta q^k)}{\partial q^i} =  \nabla_l \delta q_i + \nabla_i \delta q_l, \qquad \delta q_i = g_{ik} \delta q^k
\end{equation*}
which are the Killing equations as used in general relativity. Here we have introduced the Levi-Civita connection $\nabla$ belonging to $g_{kl}$ for which $\nabla_{i} g_{lk} = 0$. We refer to \cite{killingequationclassical} for more general connections between Noether symmetries and the Killing vectors of the corresponding Jacobi and/or Finsler metrics. Through the Killing equations, one can then potentially map our Noether procedure to studying the geodesics of a slowly changing spacetime metric, which, while interesting on its own, could pave the way to tighter connections with Wald's work, \cite{wald_original}. 
\item
The previous remark indeed brings the Noetherian setup discussed in the present paper closer to the Killing vector symmetry in \cite{wald_original}; see also Remark 5 in Section \ref{rem}.\\
One may wonder why, in the case of black hole thermodynamics, such a quasisymmetry as in \eqref{psi hamiltonian} can be derived. In his paper \cite{wald_original}, for motivating Eq (5) there, Wald mentions the diffeomorphism-invariance of the action. Similar ideas are expressed in the monograph \cite{compère2019advanced}.  That is important and striking to understand the next Section and the quasisymmetry \eqref{amat}.  Somehow, in general relativity (GR) there is no need to do the {\it extra} averaging (of the next section) to reach the thermodynamic scale; for the action of GR, one is immediately there due to its invariance under time scales.  It would be interesting to make this remark more rigorous and to give a neat relation between the existence of solutions to the Killing equations for having a Noether charge and the time-coordinate invariance in black hole GR.  

\item
One may wonder, also for the results in the next section, to what extent the results remain valid for driven or irreversible dynamics. The short answer is that the existence of the Noether charge is restricted to close-to-reversible dynamics.  The first part of the heat theorem is something that requires small nonequilibrium (driving) amplitudes; see \cite{ruelle,jonalasinio2023clausius, Saito_2011, prigo, heatconduction, Komatsu_2010, Bertini_2012, clausiusstationarystates, Maes_2015}. Technically speaking, it will be the quasisymmetry (as also referred to above) that will fail.

\end{enumerate}

\section{Thermodynamic averaging}\label{sec3}
The previous section already presents a sort of mechanical First Law with \textit{e.g.} \eqref{Bessel-Hagen with energy} and
\begin{equation*}
 \id E =    \frac 1{\xi}\id \cal N -  \partial_\lambda \cal L \,\id \lambda
\end{equation*}
 However, in the First Law of thermodynamics, one requires $\cal N = \cal N(E, \lambda)$, while $\partial_\lambda \cal L$ generally also depends on $q, \dot{q}$. That is why we require an extra averaging or projection on thermodynamic variables in what follows.

\subsection{General result}\label{subsection the proof}
We still consider the mechanical system described by the action $\mathcal{A}$ in \eqref{action in summary} or \eqref{hamiltonian action classical} but we take the protocol $\lambda^{(\ve)}(t) =\lambda(\ve t)$ for small rate $\ve$. The energy \eqref{energy function summary} is not conserved but still, for every fixed $\lambda(\tau), \tau=\ve t$, there is a conserved energy $E(\tau)$.  We now introduce the ensemble with averages $\langle \cdot  \rangle_{E(\tau),\lambda(\tau)}$ which we leave unspecified but the reader can think of the projection on the microcanonical ensemble or of time-averages.  We give examples in Section \ref{section examples}. For the moment we simply assume an averaging such that in the quasistatic limit, 
\begin{equation}\label{ama}
\lim_{\ve \to 0} \int_{\tau_1}^{\tau_2} \id \tau \ G(\tau)\Bigg[F\left(q\left(\frac{\tau}{\varepsilon}\right), \dot q\left(\frac{\tau}{\varepsilon}\right); \lambda(\tau) \right) - \left \langle F \right \rangle_{E(\tau), \lambda(\tau)} \Bigg] = 0
\end{equation}
for all smooth $G(\tau)$ and $F(t)$ (but we will need \eqref{ama} only for $F=\partial \cal L/\partial \lambda, \partial \cal H/\partial \lambda$).  In that way, ``fast'' observables $F = F(q(t),\dot{q}(t), \lambda(\tau))$ are projected on functions of the ``slow'' variables $E(\ve t)$ and $\lambda(\ve t)$.  Generally, the limit \eqref{ama} only vanishes when focusing on `thermodynamic consistent trajectories', where we borrow the terminology from \cite{Sasa_original}, {\it i.e.}, trajectories of the form
\begin{equation}\label{quasistatic classical mechanics}
    q(t) = q^*_{\lambda(\ve t)}(t) + O(\ve)
\end{equation}
which, for $\ve$ small enough, are arbitrarily close in $C^1$-norm, defined as
\begin{equation*}
    || f||_{C^1} = \sup_{t \in [t_1, t_2]} | f(t)| +  \sup_{t \in [t_1, t_2]} | f'(t)|,
\end{equation*}
to $q^*_\lambda(t)$ which is the solution of the equations of motion $\mathscr{E} =0$ at constant $\lambda$, \cite{arnold}.\\  

We take the transformation \eqref{general lagrangian transformation} with $\delta q = 0$ and where $\xi = \xi(E,\lambda)$ is a function of the energy and the protocol:
\begin{equation}\label{time translation general 2}
        t \longrightarrow t + \eta\, \xi\left(E(q_t, \dot{q}_t, \lambda^{(\ve)}_t), \lambda^{(\ve)}_t \right)
    \end{equation}
    which already appeared in \cite{Sasa_original}. Since $\lambda_t^{(\ve)} = \lambda(\ve t)$ and $E = E(\ve t)$, it follows that also $\xi = \xi(\ve t)$. This setup leads to the following central result: 

\begin{theorem}[{\bf Clausius \married \ Noether}]\label{cenr}
Suppose that the continuous symmetry \eqref{time translation general 2}
      leaves the action invariant for the class of trajectories \eqref{quasistatic classical mechanics} in the quasistatic limit $\ve \to 0$ in the sense of a quasisymmetry
    \begin{equation}\label{amat}
       \lim_{\ve \to 0} \delta \mathcal{A} = \eta \int_{\tau_1}^{\tau_2} \mathrm{d} \tau
       \frac{\mathrm{d} \Psi}{\mathrm{d} \tau}
    \end{equation}
    for some function $\Psi = \Psi(E(\tau),\lambda(\tau))$. Then, there is a state function $\cal N$ (Noether charge) so that
    \begin{equation}\label{first law theorem 2}
 \xi \,(\mathrm{d} E +  \left\langle \partial_\lambda \cal L \right\rangle_{E,\lambda}\ \mathrm{d} \lambda)
 =     \mathrm{d} (\Psi + \xi E)  =    \mathrm{d} \cal N
    \end{equation}
   is an exact differential. Moreover, on shell, $\cal N$ is conserved, {\it i.e.}, $ \mathrm{d} \cal N/\mathrm{d} \tau = 0$ along the trajectories \eqref{quasistatic classical mechanics}.
\end{theorem}

It is natural to introduce a (non-constant) temperature parameter $\beta^{-1}= k_B T= k_B T(E, \lambda)$ in the system, {\it e.g.}  $k_B T = \langle p \dot{q} \rangle_{\text{period}}$ for one-dimensional periodic systems, and we require the symmetry \eqref{time translation general 2} to hold for
\begin{equation}\label{xi as temperature}
      \xi(E, \lambda) = \hbar\, \beta(E, \lambda) = \frac{\hbar}{k_B T(E, \lambda)}
\end{equation}
where Planck's constant $\hbar$ is (only) introduced to give $\xi$ the units of time.  Then, \eqref{first law theorem 2} implies that
\begin{equation*}
    T^{-1}(E, \lambda) \,\big(  \id E + \langle \partial_\lambda \cal L \rangle_{E, \lambda} \id \lambda \big)  =\id \left(  \frac{k_B}{\hbar}  \cal N \right)
\end{equation*}
is an exact differential. We recognize $\id E + \langle \partial_\lambda \cal L \rangle_{E, \lambda} \id \lambda = \delta Q$ as the (thermodynamic) heat to the system and  $\delta W = \langle \partial_\lambda \cal L \rangle_{E, \lambda} \id \lambda = - \langle \partial_\lambda \cal H \rangle_{E, \lambda}$ is the work done on the system. Therefore, we obtain the first part of the Clausius' heat theorem \eqref{ds} which identifies the entropy $S = S(E, \lambda)$ as proportional to the Noether charge:
\begin{equation*}
    \cal N = \Psi + \xi E = \frac{\hbar S}{k_B} 
\end{equation*}
This result leads to the slogan ``Entropy as a Noether charge''. That Noether charge, and hence the entropy $S$, is an adiabatic invariant (in the thermodynamic sense) upon interpreting $T^{-1}(E, \lambda) \,\big(  \id E + \langle \partial_\lambda \cal L \rangle_{E, \lambda} \id \lambda \big)$ as thermodynamic heat.\\

We start with the proof of Theorem \ref{cenr} in the Lagrangian formalism. The line of arguing is very similar to the arguments in \cite{Sasa_original}.  The calculations have analogues in the Hamiltonian setting, which are presented in Section \ref{section hamiltonian version}. 

\begin{proof}
 Under a non-uniform time translation, {\it i.e.}, $\delta q = 0$ in \eqref{general lagrangian transformation}, the action changes as \eqref{change action lagrangian general}
\begin{align}
    & \delta \mathcal{A} 
    = \eta \int_{t_1}^{t_2} \id t \Big[- \mathscr{E} \dot{q}\, \xi - \frac{\id}{\id t} \Big( \xi\, E \Big) \Big] \label{delta A 1}
\end{align} 
We take the shift $\xi = \xi(E(q, \dot{q}, \lambda), \lambda)$ as in \eqref{time translation general 2}. The Euler-Lagrange functional $\mathscr{E}$ can be replaced from the identity \eqref{firsteuler}, such that \eqref{delta A 1} becomes
\begin{align}
    \delta \mathcal{A} &= \eta \int_{t_1}^{t_2}  \id t \Bigg[\xi(\varepsilon t) \frac{\id E}{\id t}(\varepsilon t) + \frac{\partial \cal L}{\partial \lambda}(t) \frac{\id \lambda}{\id t}(\varepsilon t) \xi(\varepsilon t)  - \frac{\id}{\id t} \Big( \xi E \Big) \Bigg] \label{change action before averaging} \\
    & = \eta \int_{\tau_1}^{\tau_2}  \id \tau \Bigg[\xi(\tau) \frac{\id E}{\id \tau}(\tau) + \frac{\partial \cal L}{\partial \lambda}\left(\frac{\tau}{\varepsilon}\right) \frac{\id \lambda}{\id \tau}(\tau) \xi(\tau)  - \frac{\id}{\id \tau} \Big( \xi E \Big) \Bigg] \nonumber
\end{align}
In the above, all functions are slowly varying, except for the $\partial \cal L/\partial \lambda$ term. Focusing (only) on thermodynamically consistent trajectories \eqref{quasistatic classical mechanics}, we use the averaging hypothesises \eqref{ama} to get
\begin{equation}\label{averaging lagrangian}
\lim_{\ve \to 0} \int_{\tau_1}^{\tau_2} \id \tau \ \frac{\id \lambda}{\id \tau}(\tau) \xi(\tau)  \Bigg[\frac{\partial \cal L}{\partial \lambda}\left(\frac{\tau}{\varepsilon}\right) - \left \langle \frac{\partial \cal L}{\partial \lambda} \right \rangle_{E(\tau), \lambda(\tau)} \Bigg] = 0
\end{equation}
and
\begin{align}
\lim_{\ve \to 0} \delta \mathcal{A} & = \eta \int_{\tau_1}^{\tau_2} \id \tau \Bigg[\xi(\tau) \frac{\id E}{\id \tau}(\tau) + \left \langle \frac{\partial \cal L}{\partial \lambda} \right \rangle_{E(\tau), \lambda(\tau)} \frac{\id \lambda}{\id \tau}(\tau) \xi(\tau)  - \frac{\id}{\id \tau} \Big( \xi E \Big) \Bigg] \label{de}
\end{align}
By assumption
\begin{equation}\label{quasisymmetry in general proof}
  \lim_{\ve \to 0}  \delta \mathcal{A} = \eta \int_{\tau_1}^{\tau_2} \id \tau \frac{\id \Psi}{\id \tau}
\end{equation}
and since (\ref{de})--(\ref{quasisymmetry in general proof}) are equal for all $\tau_1 < \tau_2$, it follows that the Noether charge $\cal N = \Psi + \xi E$ from \eqref{general noether charge lagrangian} satisfies
\begin{equation}\label{Killing equation inside the proof}
    \frac{\id \cal N}{\id \tau} = \xi \frac{\id E}{\id \tau} + \xi \left \langle \frac{\partial \cal L}{\partial \lambda} \right \rangle_{E, \lambda} \frac{\id \lambda}{\id \tau}
\end{equation}
which is \eqref{Bessel-Hagen with energy} upon averaging.
Consequently, in the quasistatic limit $\ve \to 0$, we obtain a total differential
\begin{equation*}
    \id \cal N = \xi\,\left(  \id E + \langle \partial_\lambda \cal L \rangle_{E, \lambda} \id \lambda \right)
\end{equation*}
where $\cal N$ is conserved \textit{on shell}, $\id \cal N/\id \tau = 0$. That finishes the proof of \eqref{first law theorem 2} in the Lagrangian formalism for a (non-uniform) time translation.
\end{proof}

\subsection{Remarks II}\label{rem}
\begin{enumerate}
    \item 
Continuing the first two remarks of Section \ref{rem1}, again, the above assumes the quasisymmetry \eqref{amat} of the action under the inhomogeneous time-translation \eqref{time translation general 2}.   Obviously, that assumption needs to be checked in every case.  That means that we have to look for a pair of functions $(\xi, \Psi)$ satisfying \eqref{Killing equation inside the proof}, {\it i.e.},
\begin{equation}\label{Killing equations}
    \xi(\tau) \frac{\id E}{\id \tau}(\tau) + \xi(\tau) \left \langle \frac{\partial \cal L}{\partial \lambda}\left(\frac{\tau}{\varepsilon}\right) \right \rangle_{E(\tau) , \lambda(\tau)} \frac{\id \lambda}{\id \tau}
    = \frac{\id}{\id \tau} \left(\Psi + \xi E \right)
\end{equation}
which is the (generalized) Killing equation \cite{orbits, killingequationclassical} for the transformation \eqref{time translation general 2}, expressed in terms of the energy function $E$. Alternatively, \eqref{Killing equations} can also be solved for the pair $(\xi, \cal N)$ with $\cal N = \Psi + \xi E$. We refer to Section \ref{section examples} for two clear examples, also illustrating the averaging procedure \eqref{averaging lagrangian}.  

\item  Since the Noether charge $\cal N$ is intimately related to entropy,  we expect from thermodynamics that it is only conserved under quasistatic and adiabatic conditions. To see how these considerations appear in the above derivation, note that \eqref{Killing equations} implies
\begin{align*}
    \cal N(t_2) - \cal N(t_1) &=  \int_{t_1}^{t_2} \id t \frac{\id \cal N}{\id t} = \int_{t_1}^{t_2} \id t \Bigg[\xi \frac{\id E}{\id t} + \xi \left\langle \frac{\partial \cal L}{\partial \lambda} \right\rangle \frac{\id \lambda}{\id t}\Bigg] \\
   & = \int_{t_1}^{t_2} \id t \Bigg[ -\xi \mathscr{E} \dot{q} + \xi \frac{\id \lambda}{\id t} \Bigg(  \left\langle \frac{\partial \cal L}{\partial \lambda} \right\rangle - \frac{\partial \cal L}{\partial \lambda} \Bigg)
   \Bigg]
\end{align*}
where we used \eqref{firsteuler} in the last line. This way of writing indicates that there are two sources of ``entropy'' production,
\begin{equation}\label{entropy production terms}
   \frac{\id \lambda}{\id t} \Bigg(  \left\langle \frac{\partial \cal L}{\partial \lambda} \right\rangle - \frac{\partial \cal L}{\partial \lambda} \Bigg)
   \qquad \text{ and } \qquad \mathscr{E} \dot{q}
\end{equation}
The first averaging term indicates that the entropy will only be conserved in the quasistatic limit $\ve \to 0$ where \eqref{ama} holds. The second term yields extra entropy for trajectories not on shell or when there are nonconservative forces (like friction and driving) in the Euler-Lagrange equations,  $\mathscr{E} \neq 0$. That is connected to heat through \eqref{mechanical first law lagrangian}, such that $\mathscr{E} \dot{q} = 0$ corresponds to adiabatic conditions. \\
 In general, it is not possible to derive the increase in entropy  from microscopic dynamical considerations only; in the end, the Second Law is a statistical principle which gets its stonewall character for macroscopic systems. Indeed, it is unclear how to deduce anything about the signs of the entropy production terms \eqref{entropy production terms} for a general Lagrangian system. 

\item For the averaging $\langle \cdot \rangle_{E, \lambda}$ in \eqref{ama}, we can certainly take the microcanonical ensemble $\langle \cdot \rangle^{\text{mc}}$ as was done in \cite{Sasa_original}, and the volume entropy appears,
\begin{equation}\label{volen}
    S(E, \lambda) = \log \left( \frac{\Omega(E, \lambda)}{N!}\right), \qquad \Omega(E, \lambda) = \int \id \Gamma \ \Theta\left(E- \cal H(\Gamma, \lambda) \right), \qquad \Gamma = (q,p)
\end{equation}
For time-periodic systems in one dimension, we take the Helmholtz entropy as explained in the Examples of Section \ref{section examples}.  For the general Theorem \ref{cenr} it does not matter so much as we do not concentrate on the entropy function but on characterizing ``heat over temperature'' as an exact differential.
That is somewhat different from \cite{Sasa_original} and is closer to Wald's work where the entropy is defined through the First Law of black hole thermodynamics.

\item
One may wonder about the physical interpretation of the nonuniform time translation \eqref{time translation general 2}. As was already noted in \cite{Sasa_original}, the choice $\xi = \hbar \beta$ in \eqref{xi as temperature} leads to a time translation $t \to t + \eta \hbar \beta$ which seems related to the complex time $t + i \hbar \beta$ that appears in quantum dynamics at finite temperature. See the next subsection for a different but equivalent symmetry transformation and its interpretation. \\

\item  Eq.~\eqref{first law theorem 2} is the analogue of the First Law of black hole thermodynamics; cf.~\cite{wald_original}.  
Interestingly, the appropriate symmetry  for  the Schwarzschild spacetime becomes a  field variation  along the Killing vector $\xi^\mu = (\partial/\partial t)^\mu = (1,0,0,0)$ and $\xi_\mu = \left(- (1-2 M/r),0,0,0 \right)$,  {\it i.e.}, $\delta \phi = \mathcal{L}_\xi \phi$ for all dynamical fields $\phi$ with $\mathcal{L}_\xi$ the Lie-derivative along $\xi$.  This Killing vector is related to the Hawking temperature $T_{H}$ through the relation
\begin{equation*}
    \xi^\mu \nabla_\nu \xi^\nu = \kappa \xi^\nu, \qquad T_H = \frac{\hbar c \kappa}{2 \pi k_B}
\end{equation*}
with $\kappa$ the surface gravity.  Consequently, time shifts become connected to temperature. 
\\In the following section, we show that, instead of the time-translation \eqref{time translation general 2}, one can also consider a `field variation' $q(t) \to q(t) + \eta \delta q(q(t), \dot{q}(t),t)$, which comes closer to Wald's setup. 

\end{enumerate}

\subsection{Different transformations }\label{section interpretation}

\subsubsection{Hamiltonian formalism}\label{section hamiltonian version}
\textbf{Time--shift:} \\
As before, the action changes as \eqref{change action hamiltonian general} under the non-uniform time translation \eqref{general lagrangian transformation}
\begin{align}
    \delta \mathcal{A} 
    &= \eta \int_{t_1}^{t_2} \id t \Big[- \xi(t) \dot{p}(t) \left(\dot{q} - \frac{\partial \cal H}{\partial p}  \right) + \xi(t) \dot{q}(t) \, \left(\dot{p} + \frac{\partial \cal H}{\partial q} \right) - \frac{\id}{\id t} \Big( \xi \cal H(p,q,t) \Big) \Big]  \nonumber \\
    & = \eta \int_{t_1}^{t_2} \id t \Big[ \xi(t) \dot{p}(t) \frac{\partial \cal H}{\partial p} + \xi(t) \dot{q}(t) \, \frac{\partial \cal H}{\partial q} - \frac{\id}{\id t} \Big( \xi \cal H(p,q,t) \Big) \Big] \label{delta A ham}
\end{align}
We now take the shift $\xi = \xi(\cal H(q, p, \lambda), \lambda)$ following \eqref{time translation general 2}.
As $\lambda = \lambda(\varepsilon t)$, one has $\cal H = \cal H(\varepsilon t)$ since the energy is conserved for constant $\lambda$ and thus also $\xi = \xi(\ve t)$. Using the identity \eqref{dhdt} to introduce $\id \cal H/\id t$, \eqref{delta A ham} becomes
\begin{align}
    \delta \mathcal{A} &= \eta \int_{t_1}^{t_2}  \id t \Bigg[\xi(\varepsilon t) \frac{\id \cal H}{\id t}(\varepsilon t) - \frac{\partial \cal H}{\partial \lambda}(t) \frac{\id \lambda}{\id t}(\varepsilon t) \xi(\varepsilon t)  - \frac{\id}{\id t} \Big( \xi \cal H \Big) \Bigg] \nonumber \\
    & = \eta \int_{\tau_1}^{\tau_2}  \id \tau \Bigg[\xi(\tau) \frac{\id \cal H}{\id \tau}(\tau) - \frac{\partial \cal H}{\partial \lambda}\left(\frac{\tau}{\varepsilon}\right) \frac{\id \lambda}{\id \tau}(\tau) \xi(\tau)  - \frac{\id}{\id \tau} \Big( \xi \cal H \Big) \Bigg] \label{delta a xi shift appendix}
\end{align}
In the above, all functions are slowly varying, except for the $\partial \cal H/\partial \lambda$ term. Focussing on thermodynamically consistent trajectories, we use the averaging hypothesizes \eqref{ama} to get
\begin{equation*}
\lim_{\ve \to 0}\int_{\tau_1}^{\tau_2} \id \tau \ \frac{\id \lambda}{\id \tau}(\tau) \xi(\tau)  \Bigg[\frac{\partial \cal H}{\partial \lambda}\left(\frac{\tau}{\varepsilon}\right) - \left \langle \frac{\partial \cal H}{\partial \lambda} \right \rangle_{\cal H(\tau), \lambda(\tau)} \Bigg] =0
\end{equation*}
and
\begin{align*}
 \lim_{\ve \to 0} \delta \mathcal{A} & = \eta \int_{\tau_1}^{\tau_2} \id \tau \Bigg[\xi(\tau) \frac{\id \cal H}{\id \tau}(\tau) - \left \langle \frac{\partial \cal H}{\partial \lambda} \right \rangle_{\cal H(\tau), \lambda(\tau)} \frac{\id \lambda}{\id \tau}(\tau) \xi(\tau)  - \frac{\id}{\id \tau} \Big( \xi \cal H \Big) \Bigg]
\end{align*}
By assumption
\begin{equation*}
    \lim_{\ve \to 0} \delta \mathcal{A} = \eta \int_{\tau_1}^{\tau_2} \id \tau \frac{\id \Psi}{\id \tau}
\end{equation*}
such that in the quasistatic limit, the Noether charge $\cal N = \Psi + \xi E$ from \eqref{general noether charge lagrangian} must be an exact differential, {\it i.e.},
\begin{equation}\label{dN noether}
    \id \cal N = \xi\,\left(  \id \cal H - \langle \partial_\lambda \cal H \rangle_{\cal H, \lambda} \id \lambda \right)
\end{equation}
Moreover, working on shell, $\cal N$ is conserved $\id \cal N/\id \tau = 0$.  That finishes the proof of \eqref{first law theorem 2} for the Hamiltonian case which agrees with the Lagrangian formalism since $\partial_\lambda \cal L = - \partial_\lambda \cal H$. \\

\textbf{Combined \texorpdfstring{$q,p$}{q,p}-shift:}\\
As proven in \cite{laghamconnection}, for every symmetry transformation \eqref{general hamiltonian transformation} in the Hamiltonian formalism, there is an equivalent transformation leading to the same Noether charge which uses the Poisson bracket
\begin{equation}\label{poisson bracket}
    \{F, G\} = \frac{\partial F}{\partial z} \frac{\partial G}{\partial f} - \frac{\partial G}{\partial z} \frac{\partial F}{\partial f}
\end{equation}
Briefly, let $F(q,p,t)$ be a conserved Noether charge generated by \eqref{general hamiltonian transformation}
satisfying
\begin{equation}\label{conserved quantity hamiltonian}
    \frac{\id F}{\id t} = \frac{\partial F}{\partial t} + \{F, \cal H\} = 0
\end{equation}
Then, $F$ is also the Noether charge corresponding to the transformation
\begin{align}\label{transformation appendix N}
 & t \to t' = t, \qquad  q(t) \to q'(t') = q(t) - \eta \frac{\partial F}{\partial p}, \qquad p(t) \to p'(t') = p(t) + \eta \frac{\partial F}{\partial q}
\end{align}
since, using \eqref{change action hamiltonian general} with $\xi = 0$,
\begin{align*}
      \delta \mathcal{A} &= \eta \int_{t_1}^{t_2} \id t \ \left[ \frac{\partial F}{\partial q} \left(\dot{q} - \frac{\partial \cal H}{\partial p} \right) + \frac{\partial F}{\partial p} \left( \dot{p} + \frac{\partial \cal H}{\partial q} \right) + \frac{\id}{\id t} \Big(- p \frac{\partial F}{\partial p} \Big) \right] \nonumber \\
    & =  \eta \int_{t_1}^{t_2} \id t \ \left[ \frac{\id F}{\id t} - \frac{\partial F}{\partial t} - \{F, \cal H \} + \frac{\id}{\id t} \Big( -p \frac{\partial F}{\partial p} \Big) \right] \\
    & =  \eta \int_{t_1}^{t_2}  \id t \ 
\frac{\id \Psi}{\id t}, \qquad \Psi = F -p \frac{\partial F}{\partial p}  
\end{align*}
where we have used \eqref{conserved quantity hamiltonian} in the last line. As such, the transformation \eqref{transformation appendix N} leaves th+e action invariant in the sense of a quasisymmetry \eqref{psi hamiltonian}, with corresponding Noether charge \eqref{general noether charge hamiltonian}
\begin{equation}\label{noether charge = f}
    \cal N = \Psi - p \delta q =  \Psi + p \frac{\partial F}{\partial p} =  F
\end{equation}
as desired. \\
As we mix the $q,p$ variables in \eqref{transformation appendix N}, the transformation can be interpreted as a `rotation' in phase space, akin to the transformation between $A^\mu$ and $C^\mu$ mentioned in the introduction. Moreover, the transformation \eqref{transformation appendix N} can also be written using the Poisson bracket \eqref{poisson bracket}
\begin{align}\label{poisson bracket symmetry classical}
 & q(t) \to q'(t) = q(t) - \eta \{q,F \}, \qquad p(t) \to p'(t) = p(t) - \eta \{p, F \}
\end{align}
indicating that a constant of motion generates its own symmetry. For instance, the conservation of momentum and energy are obtained from
\begin{align*}
   \text{(Momentum) :} \ & t \to t' = t, \qquad  q(t) \to q'(t') = q(t) - \eta \frac{\partial p}{\partial p} = q(t) - \eta , \qquad p(t) \to p'(t') = p(t) + \eta \frac{\partial p}{\partial q} = p(t) \\
   \text{(Energy) :} \ & t \to t' = t, \qquad  q(t) \to q'(t') = q(t) - \eta \frac{\partial \cal H}{\partial p}, \qquad p(t) \to p'(t') = p(t) + \eta \frac{\partial \cal H}{\partial q}
\end{align*}
where, most importantly, the last transformation differs from a constant translation of time. That result invites for a derivation of \eqref{dN noether} through the transformation
\begin{align}\label{q,p shift}
  &t \to t' = t, \qquad  q(t) \to  q'(t')= q(t) - \eta \frac{\partial F}{\partial p}, \qquad p(t) \to p'(t') = p(t) + \eta  \frac{\partial F}{\partial q}
\end{align}
with $F =F(\cal H(q,p, \lambda), \lambda) $. As $\lambda = \lambda(\varepsilon t)$ and $\cal H = \cal H(\varepsilon t)$ it follows that also $F = F(\ve t)$. Using $\{F, \cal H\} = 0$, the action changes as
\begin{align}
    \delta \mathcal{A} & = \eta \int_{t_1}^{t_2} \id t \ \left[ \frac{\partial F}{\partial \mathcal{H}} \left(\frac{\partial \cal H}{\partial q} \dot{q} + \frac{\partial \cal H}{\partial p} \dot{p} \right) + \frac{\id}{\id t} \Big( -p \frac{\partial F}{\partial p} \Big) \right] \nonumber \\
    &= \eta \int_{t_1}^{t_2}  \id t \ \Bigg[\frac{\partial F}{\partial \mathcal{H}} (\varepsilon t) \frac{\id \cal H}{\id t}(\varepsilon t) - \frac{\partial \cal H}{\partial \lambda}(t) \frac{\id \lambda}{\id t}(\varepsilon t) \frac{\partial F}{\partial \mathcal{H}} (\varepsilon t)  - \frac{\id}{\id t} \Big(  -p \frac{\partial F}{\partial p} \Big) \Bigg] \nonumber \\
    & = \eta \int_{\tau_1}^{\tau_2}  \id \tau \ \Bigg[ \frac{\partial F}{\partial \cal H} (\tau) \frac{\id \cal H}{\id \tau}(\tau) + \frac{\partial \cal H}{\partial \lambda}\left(\frac{\tau}{\varepsilon}\right) \frac{\id \lambda}{\id \tau}(\tau) \frac{\partial F}{\partial \cal H} (\tau)  + \frac{\id}{\id \tau} \Big( -p \frac{\partial F}{\partial p} \Big) \Bigg] \label{delta a pq shift}
\end{align}
where \eqref{dhdt} enters in the second line. Equation \eqref{delta a pq shift} has the same form as \eqref{delta a xi shift appendix} upon identifying $\xi(\tau) = \partial  F/\partial \cal H(\tau)$. Following similar steps as in the previous subsection, it then follows that the Noether charge $\cal N = \Psi + p \ \partial F/\partial p$ from \eqref{noether charge = f} must be an exact differential, {\it i.e.},
\begin{equation*}
    \id \cal N = \frac{\partial F}{\partial \cal H}\,\left(  \id \cal H - \langle \partial_\lambda \cal H \rangle_{\cal H, \lambda} \id \lambda \right)
\end{equation*}
Moreover, working on shell, $\cal N$ is conserved $\id \cal N/\id \tau = 0$.  That finishes the proof of \eqref{first law theorem 2} in the Hamiltonian formalism for the new symmetry transformation \eqref{q,p shift}. \\

In conclusion, the transformations
\begin{align}
(\text{Time translation}): \ &t \to t' = t + \eta \xi, \quad q \to q'(t') = q(t), \quad p \to p'(t') = p(t) \label{equivalence time and q,p shift}\\
   (q,p \text{ shift}): \ & t \to t' = t, \quad  q(t) \to q'(t') = q(t) - \eta \{q,F \}, \quad p(t) \to p'(t') = p(t) - \eta \{p, F \} \nonumber
\end{align}
with $\xi = \xi(\cal H(q, p,\lambda),\lambda)$ and $F = F(\cal H(q, p,\lambda),\lambda)$ lead to the same Noether charge $\cal N$ (but different $\Psi$ in the quasisymmetry) satisfying \eqref{dN noether} when $\xi = \partial F/\partial \cal H$. Moreover, when there is a (non-constant) temperature parameter $\beta^{-1} = k_B T = k_B T(\cal H, \lambda)$ in the system, one naturally requires the symmetries to hold for $\xi = \partial F/\partial \cal H = \hbar \beta$ since then \eqref{dN noether} yields
\begin{equation*}
       T^{-1}\,\left(  \id \cal H - \langle \partial_\lambda \cal H \rangle_{\cal H, \lambda} \id \lambda \right) = \id \left( \frac{k_B}{\hbar} \cal N \right) = \id S
\end{equation*}
relating the entropy to the Noether charge. The relation $ \partial F/\partial \cal H = \hbar \beta$ implies that in fact $F = \hbar S/k_B = \cal N$ which is \eqref{noether charge = f}.

\subsubsection{Lagrangian formalism}
In \cite{laghamconnection}, it was shown that there exists a similar equivalence between transformations as in \eqref{equivalence time and q,p shift} for the Lagrangian formalism, by taking the identification $p \to \partial \cal L/\partial \dot{q}$ in the symmetry transformation. That result leads to a $q$ shift of the form 
\begin{equation}\label{q shift l}
    t \to t' = t, \qquad  q(t) \to  q'(t')= q(t) - \eta \frac{\partial F}{\partial p} \Big|_{p \to \frac{\partial \cal L}{\partial \dot{q}}} = q(t) - \eta \frac{\partial F}{\partial \left( \frac{\partial \cal L}{\partial \dot{q}}\right) } = q(t) - \eta \dot{q} \frac{\partial F}{\partial E}
\end{equation}
with $F = F(E(q, \dot{q}, \lambda), \lambda) = F \left( \frac{\partial \cal L}{\partial \dot{q}} \dot{q} - \cal L, \lambda \right) $. This Lagrangian transformation can be seen as a projection in $p$ of the higher dimensional $q,p$ shift (or rotation) \eqref{q,p shift} in phase space, which was alluded to in the introduction.  We refer to Appendix \ref{appendix quasisymmetry} for a concrete example of this symmetry related to energy conservation.  \\
Under this $q$ shift \eqref{q shift l} the action changes as \eqref{change action lagrangian general},
\begin{align*}
    \delta \cal A &= \eta \int_{t_1}^{t_2} \id t \left[\mathscr{E} \delta q + \frac{\id }{\id t} \left( \frac{\partial \cal L}{\partial \dot{q}} \delta q \right) \right] \\
    & = \eta \int_{t_1}^{t_2} \id t \left[-\mathscr{E} \dot{q} \frac{\partial F}{\partial E} - \frac{\id }{\id t} \left( \frac{\partial \cal L}{\partial \dot{q}} \dot{q} \frac{\partial F}{\partial E} \right) \right] \\
    & = \eta \int_{\tau_1}^{\tau_2} \id \tau \left[ \frac{\id E}{\id \tau} \frac{\partial F}{\partial E} + \frac{\partial \cal L}{\partial \lambda}\left( \frac{\tau}{\ve} \right) \frac{\id \lambda}{\id \tau} \frac{\partial  F}{\partial E} - \frac{\id }{\id t} \left( \frac{\partial \cal L}{\partial \dot{q}} \dot{q} \frac{\partial F}{\partial E} \right) \right]
\end{align*}
where we have used the identity \eqref{firsteuler} in the last line and went over to slow time $\tau$. This result has the same form as \eqref{change action before averaging} upon identifying $\xi = \partial F/\partial E$. Following similar steps as before, one finds that the Noether charge $\cal N = \Psi - \frac{\partial \cal L}{\partial \dot{q}} \delta q =  \Psi + \frac{\partial \cal L}{\partial \dot{q}} \dot{q} \ \frac{\partial F}{\partial E}$ must be an exact differential, {\it i.e.},
\begin{equation*}
    \id \cal N = \frac{\partial F}{\partial E}\,\left(  \id E + \langle \partial_\lambda \cal L \rangle_{ E, \lambda} \id \lambda \right)
\end{equation*}
Moreover, working on shell, $\cal N$ is conserved $\id \cal N/\id \tau = 0$. That finishes the proof of \eqref{first law theorem 2} in the Lagrangian formalism for a $q$ shift, analogue to the symmetry \eqref{transformation appendix N} in the Hamiltonian formalism.

\section{Example: thermodynamic formalism of time-periodic systems}\label{section examples}

For pedagogical purposes, we end by making contact with celebrated mechanical examples that were central in the pioneering attempts to give mechanical foundations to thermodynamics.  They are the one-dimensional periodic systems with period $\mathcal{T}$, where the averaging $\langle \cdot \rangle_{E, \lambda}$ in \eqref{ama} was understood as time-averaging, {\it i.e.},
\begin{equation}\label{time average}
    \langle f \rangle_{\mathcal{T}_{\lambda(\tau)}} = \frac{1}{\mathcal{T}_{\lambda(\tau)}} \int_0^{\mathcal{T}_{\lambda(\tau)}} \id t \ f(t; \lambda(\tau)), \qquad \mathcal{T} = \mathcal{T}(E, \lambda)
\end{equation}
Importantly, in this average one keeps $\lambda(\tau)$  constant in the integrand and we only integrate out the fast variables. 
We refer to \cite{Gal2,thermo_kepler} for the history, the relation with ergodicity and the further developments into a chaotic hypothesis.\\
In that respect we need to recall the Helmholtz entropy $S_\text{H}$,  \cite{Helmholtza, Helmholtzb},  for one-dimensional time-periodic systems with Hamiltonian $\mathcal{H}(p,q;\lambda) = p^2/(2m) + V(q,\lambda)$, 
\begin{equation}\label{helmh}
S_\text{H}(E,\lambda) = \log \left( 2\int_{x_-(E,\lambda)}^{x_+(E,\lambda)} \frac{\id q}{2 \pi \hbar} \,\sqrt{2m(E-V(q,\lambda))} \right) = \log \left( \oint \frac{\id q}{2 \pi \hbar} \ p(q) \right) = \log \left( \frac{I}{\hbar} \right)
\end{equation}
with $I$ the adiabatic invariant. In fact, that Helmholtz entropy $S_\text{H}$ is then (for one-dimensional periodic systems) equal to the volume entropy \eqref{volen}. Hence, $S_\text{H}$ represents a (mechanical) one-dimensional analogue of the Clausius thermodynamic entropy:
\begin{equation*}
    \id S_\text{H} = \frac{\id E + F\id\lambda}{T}
\end{equation*}
when $T$ is twice the time-average of $K=p^2/(2m)$ and $F$ is minus the time-average of $\frac{\partial V}{\partial \lambda}$.

\subsection{Harmonic oscillator}
If only for pedagogical purposes, we consider a classical  one-dimensional harmonic oscillator with slowly varying frequency $\omega(\ve t)$, with Hamiltonian $\cal H$ and Lagrangian $\cal L$
\begin{align}
   & \cal H(q,p,t) =  \frac{p^2}{2 m} + \frac{m \omega(\ve t)^2}{2} q^2 , \qquad \cal L(q, \dot{q},t) =  \frac{m}{2}\dot{q}^2 - \frac{m \omega(\ve t)^2}{2} q^2  \label{H and L ho}
\end{align}
The adiabatic invariant $I$ for the harmonic oscillator equals $I = E/\omega$ \cite{landau1976mechanics}, such that the Helmholtz entropy \eqref{helmh} is
\[
 S_\text{H}(E,\omega) = k_B \log \left( \frac{I}{\hbar} \right) = k_B \log \left( \frac{E}{\hbar \omega}\right)
\]
Again, we know that the Helmholtz entropy is invariant (for constant $\omega$) under adiabatic changes, a result that calls for Noether.\\

As explained in Section \ref{subsection the proof}, we need  a pair of functions $(\xi, \Psi)$ that satisfies the Killing equation \eqref{Killing equations},
\begin{equation}\label{Killing equations ho}
    \xi(\tau) \frac{\id E}{\id \tau}(\tau) + \xi(\tau) \left \langle \frac{\partial \cal L}{\partial \omega}\left(\frac{\tau}{\varepsilon}\right) \right \rangle_{E(\tau) , \omega(\tau)} \frac{\id \omega}{\id \tau} 
    = \frac{\id}{\id \tau} \left(\Psi + \xi E \right)
\end{equation}
for an appropriate averaging procedure $\langle \cdot \rangle_{E, \omega}$ that satisfies \eqref{averaging lagrangian}. Here we consider the time-average
\begin{equation*}
    \langle f \rangle_{E, \omega} = \frac{1}{\mathcal{T}}  \int_0^{\mathcal{T}} \id t \ f\left(q(t), \dot{q}(t), \omega \right), \qquad \mathcal{T} = \frac{2 \pi}{\omega} 
\end{equation*}
in which $\omega$ is kept constant as it varies slowly compared to the fast variables $q(t), \dot{q}(t)$. Consequently, for thermodynamic consistent trajectories \eqref{quasistatic classical mechanics} of the form $ q(t)  = A(\tau) \cos(\omega(\tau) t + \delta) + O(\ve)$ with $\dot{q}(t) = - A(\tau) \omega(\tau) \sin(\omega(\tau) t + \delta)  + O(\ve)$, one finds
\begin{eqnarray}
 \langle q^2 \rangle_{E(\tau), \omega(\tau)} = \frac{A(\tau)^2}{2} 
     ,&&  \langle \dot{q}^2 \rangle_{E(\tau), \omega(\tau)}  = \frac{\omega(\tau)^2 A(\tau)^2}{2} \nonumber \\
      E(\tau) &= & \langle E \rangle_{E(\tau), \omega(\tau)} = \frac{m \omega(\tau)^2 A(\tau)^2}{2} \nonumber \\
        \left \langle \frac{\partial \cal L}{\partial \omega} \right \rangle_{E(\tau), \omega(\tau)} &=& - \left \langle \frac{\partial E}{\partial \omega} \right \rangle_{E(\tau), \omega(\tau)} =
       - \frac{E(\tau)}{\omega(\tau)} \label{averaged lagrangian ho}
\end{eqnarray}
As expected, one can equivalently do a microcanonical average
\begin{align*}
    \langle q^2 \rangle^{\text{mc}} &= \frac{1}{\cal W} \int \id q \ \id p \ q^2 \,\delta(H-E(\tau))  = \frac{E(\tau)}{m \omega(\tau)^2} = \frac{A(\tau)^2}{2} \\
    {\cal W} &= \int \id q \ \id p \ \delta(H - E(\tau)) = \frac{2 \pi}{\omega(\tau)} = \mathcal{T}(\tau) 
\end{align*}
Then, specifically, we find
\begin{align*}
& \lim_{\ve \to 0} \int_{\tau_1}^{\tau_2} \id \tau \ \xi(\tau) \frac{\id \omega}{\id \tau}(\tau)  \Bigg[\frac{\partial \cal L}{\partial \omega}\left(\frac{\tau}{\varepsilon}\right) - \left \langle \frac{\partial \cal L}{\partial \omega} \right \rangle_{E(\tau), \omega(\tau)} \Bigg]  \\
    &  = - \lim_{\ve \to 0}  \int_{\tau_1}^{\tau_2} \id \tau \ \xi(\tau) \frac{\id \omega}{\id \tau} m \omega(\tau)  \Bigg[q \left(\frac{\tau}{\ve} \right)^2 - \frac{A(\tau)^2}{2} + O(\ve) \Bigg] \\
    & = -\lim_{\ve \to 0} \int_{\tau_1}^{\tau_2} \id \tau \ \xi(\tau) \frac{\id \omega}{\id \tau} m \omega(\tau) \frac{A(\tau)^2}{2}  \Bigg[ 2 \cos^2\left(\omega(\tau) \frac{\tau}{\ve} + \delta\right) - 1 + O(\ve) \Bigg] \\
    & = - \Re\left\{ \lim_{\ve \to 0} \int_{\tau_1}^{\tau_2} \id \tau \ \xi(\tau) \frac{\id \omega}{\id \tau}\frac{E(\tau)}{\omega(\tau)} \cdot  \frac{e^{2 i \left( \omega(\tau) \frac{\tau}{\ve} + \delta \right)} + e^{- 2 i \left(  \omega(\tau) \frac{\tau}{\ve} + \delta\right)}}{2} \right\} \\
    & = 0
\end{align*}
where the last integral is zero due to the rapid oscillations of the integrand, following the stationary phase approximation, \cite{mathmethods}.
Condition \eqref{averaging lagrangian} is therefore satisfied. Using \eqref{averaged lagrangian ho}, the Killing equation \eqref{Killing equations ho} then becomes
\begin{equation}\label{killing eq energy ho}
    \xi \frac{\id E}{\id \tau} - \xi \frac{E}{\omega} \frac{\id \omega}{\id \tau} = \frac{\id \cal N}{\id \tau}
\end{equation}
with $\cal N = \Psi + \xi E$. There is a natural choice for the temperature
\begin{equation}\label{beta ho}
   \beta^{-1}(E) = k_B T(E) = \langle p \dot{q} \rangle_{E, \omega} = \frac{\omega}{2 \pi} \int_0^{\frac{\omega}{2 \pi}} \id t \ p \ \dot{q} = 2 \frac{\omega}{2 \pi} \int_{- \sqrt{\frac{2 E}{m \omega^2}}}^{\sqrt{\frac{2 E}{m \omega^2}}} \id q \ \sqrt{2m E - m^2 \omega^2 q^2} = E
\end{equation}
in agreement with the equipartition theorem, such that
\begin{equation*}
    \xi(E) = \hbar \beta(E) = \frac{\hbar}{E}, \qquad  \cal N(E, \omega)  = \frac{\hbar S_\text{H}(E, \omega)}{k_B} =  \hbar  \log \left( \frac{E}{\hbar \omega} \right)
\end{equation*}
gives 
\begin{equation}
  \frac{\id \cal N}{\id \tau} = \frac{\hbar}{E} \frac{\id E}{\id \tau} - \frac{\hbar}{\omega} \frac{\id \omega}{\id \tau} =  \xi \frac{\id E}{\id \tau} - \xi \frac{E}{\omega} \frac{\id \omega}{\id \tau}
\end{equation}
which is \eqref{killing eq energy ho}. (As mentioned in Section \ref{subsection the proof}, the factor $\hbar$ is introduced to give $\xi$ the correct units of time.) 
Since $\cal N = \Psi + \xi E$ is the Noether charge \eqref{general noether charge lagrangian}, it follows that the entropy $S_\text{H}$ can be seen as the Noether invariant associated with the transformation 
\begin{equation*}
    t \to t' = t + \eta \hbar\, /E(q(t), \dot{q}(t), \omega(\ve t)) 
\end{equation*}
in the quasistatic limit $\ve \to 0$ for the subclass of thermodynamically consistent trajectories \eqref{quasistatic classical mechanics}. \\

On the other hand, taking \begin{equation*}
    \xi(\omega) = \frac{1}{\omega}, \qquad  \cal N(E, \omega)  = I(E, \omega) = \frac{E}{\omega}
\end{equation*}
yields
\begin{equation}
  \frac{\id \cal N}{\id \tau} = \frac{1}{\omega} \frac{\id E}{\id \tau} - \frac{E}{\omega^2} \frac{\id \omega}{\id \tau} =  \xi \frac{\id E}{\id \tau} - \xi \frac{E}{\omega} \frac{\id \omega}{\id \tau}
\end{equation}
which is \eqref{killing eq energy ho}. Consequently, shifting time with the (time-dependent) frequency yields the adiabatic invariant $I$ as a Noether charge in the quasistatic limit $\ve \to 0$, which agrees with \cite{orbits}.

\subsection{Kepler problem}
We consider a mass $m$ in an elliptic orbit around a larger central mass $M$ with slowly varying gravitational attraction strength $g(\ve t)$. Since $g = G M$, we imagine here a time-dependent Newton constant $G$ and/or central mass $M$. For the latter, one can think of a star losing mass through radiation ($\dot{M} < 0$) or a planet being bombarded by asteroids $(\dot{M} > 0$).  The Hamiltonian and Lagrangian are
\begin{align}
    & \mathcal{H} =    \frac{p_{r}^2}{2m} + \frac{p_{\theta}^2}{2 m r^2} - \frac{m g(\ve t)}{r}, \qquad  \mathcal{L} = \frac{m}{2}  \dot{r}^2  + \frac{m}{2} r^2 \dot{\theta}^2  + \frac{m g(\ve t)}{r}  \label{full hamiltonian kepler}
\end{align}
in the usual polar coordinates.\\
When $g$ is constant,  the elliptic orbit is characterized by angular momentum $L$ and energy $E <0$
\begin{align}
    &L =   m r_i^2 \dot{\theta}_i = p_{\theta,i} = \text{ constant},\;\quad E  = - \frac{m g}{2a}= \text{ constant} < 0 \label{l conservation kepler} \\
    & 1-e^2 = \frac{L^2}{m^2 a g} = \frac{2 |E | L^2}{m^3 g^2}, \qquad \mathcal{T}^2 = \frac{4 \pi^2}{g} a^3 \nonumber \\
    & r(t) = \frac{L^2}{m^2 g (1 + e \cos(\theta(t) - \theta_0))} = \frac{a (1-e^2)}{1 + e \cos(\theta(t) - \theta_0)} \nonumber
\end{align}
with $a$ the semi-major axis, $e$ the eccentricity and $\mathcal{T}$ the period. That allows to rewrite \eqref{full hamiltonian kepler}
\begin{align}
    & \cal H
    =    \frac{p_{r}^2}{2m} + \frac{L^2}{2 m r^2} - \frac{m g}{r}, \qquad  \cal L =  \frac{1}{2} m \dot{r}^2 - \frac{L^2}{2m r^2} + \frac{m g}{r}  \label{lagrangian kepler constant g}
\end{align}
where we  take $g \to g(\ve t)$, keeping $L,m$ constant. 
The radial coordinate $r$ is bounded for the elliptic orbit, {\it i.e.}, $r_{\text{min}} \leq r \leq  r_{\text{max}}$ where 
$r_{\text{min}}, r_{\text{max}}$ solve 
$E = -\frac{m g}{r} + \frac{L^2}{2 m r^2}$:
\begin{equation*}
     r_{\text{min}} = \frac{ m g - \sqrt{(mg)^2 -  \frac{2 |E| L^2}{m}}}{2 |E|}, \qquad r_{\text{max}} = \frac{m g + \sqrt{(mg)^2 - \frac{2 E L^2}{m}}}{2 |E|}
\end{equation*}
The adiabatic invariant for the (radial) Kepler problem corresponding to \eqref{lagrangian kepler constant g} is given by, \cite{landau1976mechanics},
\begin{equation*}
    I_r = - L + m g \sqrt{\frac{m}{2  |E|}}
\end{equation*}
 and the Helmholtz entropy \eqref{helmh} equals
\begin{equation}\label{kepler entropy}
     S_\text{H}(E,g) = k_B \log \left(\frac{I_r}{\hbar} \right) = k_B \log \left( - \frac{L}{\hbar} + \frac{ m g}{\hbar} \sqrt{\frac{m}{ 2|E|}}\right)
\end{equation}
To understand the entropy \eqref{kepler entropy} as a Noether charge, we remember to take $g = g(\ve t)$ slowly varying.
As before, we need to find a pair of functions $(\xi, \Psi)$ that satisfies the Killing equation \eqref{Killing equations},
\begin{equation}\label{Killing equations kepler}
    \xi(\tau) \frac{\id E}{\id \tau}(\tau) + \xi(\tau) \left \langle \frac{\partial \cal L}{\partial g}\left(\frac{\tau}{\varepsilon}\right) \right \rangle_{E(\tau) , g(\tau)} \frac{\id g}{\id \tau}
    = \frac{\id}{\id \tau} \left(\Psi + \xi E \right)
\end{equation}
for an appropriate averaging procedure $\langle \cdot\rangle_{E, g}$ that satisfies \eqref{averaging lagrangian}. We consider the time average
\begin{equation*}
    \left\langle f \right\rangle_{E, g} = \frac{1}{\mathcal{T}} \int_0^{\mathcal{T}} \id t \ f(t) = \frac{1}{\mathcal{T}} \int_0^{2 \pi} \id\theta \ f(\theta) \ \frac{m r(\theta)^2}{L} 
\end{equation*}
where we have used the conservation of angular momentum $L$ in \eqref{l conservation kepler}. Focussing on thermodynamic consistent trajectories \eqref{quasistatic classical mechanics},
\begin{align}
    r(t) &= \frac{L^2}{m^2 g(\ve t) (1 + e \cos(\theta(t) - \theta_0))}, \qquad \left \langle \frac{1}{r} \right \rangle_{E(\tau), g(\tau)} = \frac{1}{a(\tau)} \nonumber \\
      & \hspace{-1 cm} \left \langle \frac{\partial \cal L}{\partial g} \right \rangle_{E(\tau), g(\tau)} = - \left \langle \frac{\partial E}{\partial g} \right \rangle_{E(\tau), g(\tau)} = \frac{m}{a(\tau)} = -\frac{2 E(\tau)}{g(\tau)} \label{averaged lagrangian kepler}
\end{align}
Equivalently, in the microcanonical ensemble
\begin{align*}
    \left \langle \frac{1}{r} \right \rangle^{\text{mc}} &= \frac{1}{\cal W} \int \id r \ \id p_r \ \frac{1}{r} \,\delta(H-E(\tau))  = -\frac{2 E(\tau)}{m g(\tau)} = \frac{1}{a(\tau)} \\
    {\cal W} &= \int \id r \ \id p_r \ \delta(H - E(\tau)) = \frac{\pi g(\tau)}{\sqrt{2}} \left(\frac{m}{|E(\tau)|} \right)^{3/2}
\end{align*}
Explicitly, by again applying the stationary phase approximation, \cite{mathmethods},
\begin{align*}
    & \lim_{\ve \to 0} \int_{\tau_1}^{\tau_2} \id \tau \ \xi(\tau) \frac{\id g}{\id \tau} \Bigg[ \frac{\partial \cal L}{\partial g}\left( \frac{\tau}{\ve} \right) - \left \langle \frac{\partial \cal L}{\partial g} \right \rangle_{E(\tau), g(\tau)} \Bigg] \\
    & = \lim_{\ve \to 0} \int_{\tau_1}^{\tau_2} \id \tau \ \xi(\tau) \frac{\id g}{\id \tau} \Bigg[ \frac{m}{r(\tau/\ve)} - \frac{m}{a(\tau)}\Bigg] \\
    &  = \lim_{\ve \to 0} \int_{\tau_1}^{\tau_2} \id \tau \ \xi(\tau) \frac{\id g}{\id \tau} \frac{m \ e(\tau)}{a(\tau)} \Bigg[ \frac{\cos( \nu(\tau/\ve))}{1 - e \cos(\nu(\tau/\ve))} \Bigg] \\
    &  = \Re\left\{ \lim_{\ve \to 0} \int_{\tau_1}^{\tau_2} \id \tau \ \xi(\tau) \frac{\id g}{\id \tau} \frac{m \ e(\tau)}{a(\tau)} \Bigg[ \frac{\exp({i \nu(\tau/\ve)}) + \exp(-i \nu(\tau/\ve)) }{2 - e ( \exp({i \nu(\tau/\ve)}) + \exp(-i \nu(\tau/\ve))) } \Bigg] \right\} \\
    & = 0
\end{align*}
where we have introduced the eccentric anomaly 
$\nu(t)$ through $r(t) = a [1 - e \cos(\nu(t))]$, which satisfies Kepler's equation 
\begin{equation}\label{eccentric anomaly}
    \frac{2 \pi}{\mathcal{T}} t = \nu(t) - e \sin(\nu(t)) \Longrightarrow \nu(t) = \frac{2 \pi t}{\mathcal{T}} + \sum_{k = 1}^{\infty} \frac{2}{k} J_{k}(k e) \sin \left(\frac{2 \pi k}{\mathcal{T}} t \right) 
\end{equation}
with $J_k(x)$ the Bessel function of order $k$, \cite{goldstein2002classical}. 
We note here that the stationary phase approximation applies since $\nu(\tau/\ve) \sim \phi(\tau)/\ve $ based on \eqref{eccentric anomaly}.  Therefore, condition \eqref{averaging lagrangian} is satisfied.\\ Using \eqref{averaged lagrangian kepler}, the Killing Equation \eqref{Killing equations kepler} with $\cal N = \Psi + \xi E$, becomes
\begin{equation}\label{killing eq energy kepler}
    \xi \frac{\id E}{\id \tau} - \xi \frac{2 E}{g} \frac{dg}{\id \tau} = \frac{\id \cal N}{\id \tau}
\end{equation}
There is a natural choice for the temperature
\begin{equation*}
   k_B T(E) = \langle p_r \dot{r} \rangle_{E, g} = \frac{1}{\mathcal{T}} \int_0^{\mathcal{T}} \id t \ p_r \ \dot{r} = 2 \frac{1}{\mathcal{T}} \int_{r_{\text{min}}}^{r_{\text{max}}} \id r \ \frac{\sqrt{(r-r_{\text{min}}) (r_{\text{max}}-r)}}{r} = 2 |E| - \frac{2 \sqrt{2} |E|^{3/2} L}{m^{3/2} g}
\end{equation*}
For $L = 0$, the motion is one dimensional such that $|E| = k_B T/2$ agrees with the equipartition theorem. Consequently, 
\begin{align*}
    \xi(E,g) &= \hbar \beta(E,g) = \frac{\hbar m^{3/2} g}{2 \left(m^{3/2} g  |E| - \sqrt{2} |E|^{3/2} L \right)} \\
    \cal N(E, g) &= \frac{\hbar S_\text{H}(E,g)}{k_B} =  \hbar \log \left( - \frac{L}{\hbar} + \frac{\pi m^{3/2} g}{2 \pi \hbar} \sqrt{\frac{2}{|E|}}\right) 
\end{align*}
yields
\begin{equation*}
    \frac{\id \cal N}{\id \tau} = \frac{\hbar m^{3/2} g}{2 \left(m^{3/2} g  |E| - \sqrt{2} |E|^{3/2} L \right)} \frac{\id E}{\id \tau} + \frac{\hbar m^{3/2}}{ m^{3/2} g - \sqrt{2 |E|} L } \frac{\id g}{\id \tau} =  \xi \frac{\id E}{\id \tau} - \xi \frac{2 E}{g} \frac{dg}{\id \tau}
\end{equation*}
which is \eqref{killing eq energy kepler}. 
Since $\cal N = \Psi + \xi E$ is the Noether charge \eqref{general noether charge lagrangian}, it follows that the entropy $S_\text{H}$ can be seen as the Noether invariant associated with the transformation 
\begin{equation*}
    t \to t' = t + \eta \hbar \beta\Big(E(r(t), \dot{r}(t), g(\ve t)), g(\ve t) \Big)
\end{equation*}
in the quasistatic limit $\ve \to 0$ for the subclass of thermodynamically consistent trajectories \eqref{quasistatic classical mechanics}.

\section{Conclusion}
We have related the first part of Clausius' heat theorem with Noether's theorem, yielding an exact differential which is conserved \textit{on shell}. Our procedure focuses on deriving the First Law to identify ``heat over temperature'' as a Noether charge.  The type of entropy depends on the averaging for obtaining thermodynamically consistent trajectories.  {\it E.g.}, time-averaging brings back the early work by Clausius and Boltzmann on entropy using time-periodic one-dimensional systems.  In the case of black hole thermodynamics, no averaging appears necessary as is (we believe) a consequence of the invariance of the action under diffeomorphisms.\\
We have elaborated on the connection between the pioneering contributions in \cite{wald_original, wald2, Sasa_original,langevin_noether} via the unifying formalism of solving Killing equations and by connecting the various applied continuous symmetries.  More can be done here:
\\

It is not entirely clear how the present analysis relates to Wald's work on black hole thermodynamics. While the final result ``entropy as Noether charge'' is the same, Wald does not seem to require a quasistatic setup with parameters $\lambda(\ve t)$ (the black hole is stationary from the beginning) or an averaging procedure $\langle \cdot \rangle_{E(\tau), \lambda(\tau)}$, possibly due to the diffeomorphism invariance of the action. Clarifying that connection will lead to a deeper understanding of gravitational thermodynamics, as also the Kepler example in Section IV.B appears to indicate.\\
Furthermore, the approach can be generalized to (relativistic) field theories or hydrodynamic theories, allowing to identify entropy starting from an action principle (as envisioned also by Sasa in \cite{Sasa_original}). See also \cite{inflow, entropysuperspace, glorioso2017secondlawthermodynamicssymmetry}. In such hydrodynamic and thermodynamic contexts, there appears an interesting relation with variational methods, at least close to equilibrium, as was pursued {\it e.g.} in \cite{Giorgio1, Giorgio2,Beyengradientflow, beyengeneric}.\\

\textbf{Acknowledgment:} AB is supported by the Research Foundation - Flanders (FWO) doctoral fellowship 1152725N.

\appendix

\section{Energy conservation through a \texorpdfstring{$\delta q$}{δq} shift}\label{appendix quasisymmetry}
As mentioned in section II.B, one typically derives energy conservation when $\frac{\partial \cal L}{\partial t} = 0$ from a strict, constant time translation symmetry, i.e. $\delta q = 0, \xi = $ constant $ = \Delta t$, $\Psi = 0$, since then
\begin{align*}
    \delta \mathcal{A} &= \eta \int_{t_1}^{t_2} \id t \ \left[ - \Delta t \ \dot{q} \left(\frac{\partial \cal L}{\partial q} - \frac{\id}{\id t} \left( \frac{\partial \cal L}{\partial \dot{q}} \right) \right) \, - \Delta t \ \frac{\id}{\id t}\left(\frac{\partial \cal L}{\partial \dot{q}} \,\dot{q} - \cal L\right) \right] \\
    & = \eta \int_{t_1}^{t_2} \id t \ \Delta t \ \ \frac{\partial \cal L}{\partial t} = 0 \\
    & \Longrightarrow \cal N = \Psi + \left( \frac{\partial \cal L}{\partial \dot{q}} \dot{q} - \mathcal{L} \right) \Delta t \ = E(q, \dot{q}) \ \Delta t
\end{align*}
with energy function $E$. Differently, since we know we want to obtain $E$ as a Noether charge, we follow Section III.C and consider a $\delta q$ transformation as in \eqref{q shift l} with $F = E \Delta t$, where a factor $\Delta t$ is added to give $F$ the correct units. That transformation corresponds to  $\xi = 0, \delta q = - \Delta t \ \dot{q}$, such that the change in the action yields
\begin{align*}
     \delta \mathcal{A} &= \eta \int_{t_1}^{t_2} \id t \ \left[ - \Delta t \ \dot{q}\left(\frac{\partial \cal L}{\partial q} - \frac{\id}{\id t} \left( \frac{\partial \cal L}{\partial \dot{q}} \right) \right) \, - \frac{\id}{\id t} \left( \frac{\partial \cal L}{\partial \dot{q}} \Delta t \ \dot{q}  \right) \right] \\
     &= \eta \int_{t_1}^{t_2} \id t \ \left[ - \Delta t \ \dot{q} \frac{\partial \cal L}{\partial q}  - \frac{\partial \cal L}{\partial \dot{q}} \Delta t \ \ddot{q}  \right] \\
     &= \eta \int_{t_1}^{t_2} \id t \ \frac{\id}{\id t} \left[ -\cal{L} \Delta t \right] \\
     &= \eta \int_{t_1}^{t_2} \id t \ \frac{\id \Psi}{\id t}, \qquad \Psi = -\cal{L} \Delta t \\
     &\Longrightarrow \cal N = \Psi - \frac{\partial \cal L}{\partial \dot{q}} \delta q = \left(\frac{\partial \cal L}{\partial \dot{q}} \dot{q} - \cal L \right) \Delta t = E(q, \dot{q}) \ \Delta t
\end{align*}
which is the same as before. This example illustrates the content of Section III.C and shows that the Noether symmetry transformation of a given charge is not unique.

\bibliographystyle{unsrt}
\bibliography{bib.bib}

\end{document}